# Quantification of cardiac capillarization in single-immunostained myocardial slices using weakly supervised instance segmentation

Zhao Zhang[1,+], Xiwen Chen[2,+], William Richardson[3], Bruce Z. Gao[1], Abolfazl Razi[2] and Tong Ye[1,4,*]

[1]Department of Bioengineering, Clemson University, Clemson, SC, USA

[2]School of Computing, Clemson University, Clemson, SC, USA

[3]Department of Chemical Engineering, University of Arkansas, Fayetteville, AR, USA

[4]Department of Regenerative Medicine and Cell Biology, Medical University of South Carolina, Charleston, SC, USA

*Corresponding: ye7@clemson.edu

+These authors contributed equally to this work.

## Abstract

Decreased myocardial capillary density has been reported as an important histopathological feature associated with various heart disorders. Quantitative assessment of cardiac capillarization typically involves double immunostaining of cardiomyocytes (CMs) and capillaries in myocardial slices. In contrast, single immunostaining of basement membrane components is a straightforward approach to simultaneously label CMs and capillaries, presenting fewer challenges in background staining. However, subsequent image analysis always requires manual work in identifying and segmenting CMs and capillaries. Here, we developed an image analysis tool, AutoQC, to automatically identify and segment CMs and capillaries in immunofluorescence images of collagen type IV, a predominant basement membrane protein within the myocardium. In addition, commonly used capillarization-related measurements can be derived from segmentation masks. AutoQC features a weakly supervised instance segmentation algorithm by leveraging the power of a pre-trained segmentation model via prompt engineering. AutoQC outperformed YOLOv8-Seg, a state-of-the-art instance segmentation model, in both instance segmentation and capillarization assessment. Furthermore, the training of AutoQC required only a small dataset with bounding box annotations instead of pixel-wise annotations, leading to a reduced workload during network training. AutoQC provides an automated solution for quantifying cardiac capillarization in basement-membrane-immunostained myocardial slices, eliminating the need for manual image analysis once it is trained.

# Introduction

Capillaries in cardiac tissues play an important role in maintaining cardiac function by regulating local blood perfusion and metabolic exchange. Vascular endothelial dysfunction and cardiac capillary rarefaction, have been reported to be correlated with heart disorders, such as diabetic cardiomyopathy[1,2], dilated cardiomyopathy[3], hypertrophic cardiomyopathy[4,5], uremic cardiomyopathy[6], aging[7–9], and hypertensive heart disease[10,11]. Evaluation of cardiac capillarization typically involves microscopic examination of immunostained tissue sections. Quantitative measurements of cardiac capillarization, such as capillary density and capillary-to-cardiomyocyte ratio, can be acquired by microscopic image analysis of immunostained cardiomyocytes (CMs) and capillaries.

Immunostaining for simultaneous detection of CMs and capillaries remains problematic, complicating the quantitative capillarization assessment in myocardial slices. For example, given limited host species for antibody production (most raised in rabbits and mice), multiple antigen detection in mouse tissue can be challenging due to the high background caused by using antibodies raised in mice. Several protocols and commercial kits have been developed to suppress "mouse-on-mouse" background staining; however, their effective applications to specific immunostaining experiments require comprehensive assessment and optimization. Another solution is the direct labeling of blood vessels and cardiac sarcolemma using fluorescent-conjugated markers, such as *Griffonia Simplicifolia* isolectin B4 and wheat germ agglutinin. These markers have limited specificity, and their efficacy varies across different species and tissues[12–14]. Instead of double immunostaining, single immunostaining of basement membrane proteins provides an easy and robust way to detect CMs and capillaries simultaneously[10,15,16]. When conducting quantitative image analysis, it is essential for researchers to have expertise in identifying, counting, and tracing CMs and capillaries in single-immunostaining images. Accordingly, an automated image analysis tool for classification and segmentation of CMs and capillaries will make the single-immunostaining approach more efficient and practical for quantitative assessment of cardiac capillarization.

Deep learning (DL) approaches have been implemented to automate the classification and segmentation of biological structures in microscopic images[17–19], majorly involving semantic and instance segmentation. Instance segmentation goes beyond semantic segmentation, since its capability to identify each object allows for counting objects under specific categories. Mask R-CNN[20], PANet[21], and YOLO-series (after

YOLOv5)[22] are among the most popular instance segmentation models. These models rely on end-to-end learning from large-scale datasets for sufficient training, requiring laborious data acquisition and pixel-wise annotation. Inspired by prompt-based learning in natural language processing[23], adapting large pre-trained vision models to downstream tasks via prompt engineering has emerged as a prominent approach to achieve few- or zero-shot transfer, i.e., adapting to new scenarios with few or no labeled data[24]. Recently, a pre-trained image segmentation model, Segment Anything Model (SAM)[25], has been demonstrated with impressive zero-shot transfer capability to new datasets via prompt engineering. The implementation of engineered prompts on SAM provides an opportunity to solve segmentation problems on custom datasets, such as microscopic bioimages. However, SAM lacks classification capability; additional networks are required if classification is needed in image analysis.

Here, we propose a DL approach, named AutoQC, for automated identification and segmentation of CMs and capillaries in immunofluorescence images, enabling efficient quantification of cardiac capillarization in single-immunostained myocardial slices. AutoQC features a weakly supervised instance segmentation model by integrating a SAM-based segmentation block with a custom-designed prompt-learning block. The integration of the prompt-learning block effectively adapts SAM for microscopic bioimage datasets and enables AutoQC to classify objects. Furthermore, common capillarization-related measurements can be automatically computed and output by AutoQC. An ablation study was conducted to evaluate the efficacy of the prompt design in AutoQC. The performance of AutoQC in instance segmentation and capillarization assessment was compared with the SAM without prompting (SAM-Only) and a state-of-the-art instance segmentation model, YOLOv8-Seg.

## Results

**Ablation study**

We conducted an ablation study to demonstrate the effectiveness of our prompt-design approach. The model that SAM was prompted by sets of binary-labeled point only was named P-SAM, while the model that SAM was prompted by bounding boxes only was named BB-SAM (see **Design of the prompt-learning block**). Both P-SAM and BB-SAM utilized the same object detector as AutoQC. Performance evaluation of instance segmentation was conducted among P-SAM, BB-SAM, and AutoQC.

Figure 1 shows representative segmentation masks predicted by P-SAM, BB-SAM and AutoQC. P-SAM showed limited capability of handling densely packed objects, while BB-SAM had improved performance in object localization and boundary detection. AutoQC with enhanced prompts, the combination of keypoints and bounding boxes, demonstrated better segmentation performance than P-SAM and BB-SAM. Since all models shared the same object detector, there was no difference in object classification (CM or capillary). Figure 2 shows segmentation evaluation results of five test runs ($n = 5$). For each test run, mAP, mAR and F1 score were computed across the test dataset ($N = 32$). BB-SAM outperformed P-SAM in terms of mAP, mAR, and F1 score. AutoQC achieved significantly higher mAP, mAR, and F1 score compared to P-SAM (all $p$ values < 0.0001) and BB-SAM (all $p$ values < 0.05). In addition, AutoQC demonstrated the highest mAP, mAR, and F1 score (0.824, 0.844, and 0.834) when computed using single-value IoU threshold, 0.5. The evaluation results obtained from five test runs are reported as mean ± SD in Table 1.

**Benchmarking**

AutoQC was benchmarked against YOLOv8-Seg, a state-of-the-art end-to-end model for instance segmentation. YOLOv8-Seg was trained by pixel-wise annotated dataset (see **Image acquisition and dataset preparation**). Additionally, YOLOv8-Seg utilized the same model setup as the YOLO-based object detector in AutoQC (see **Model setup**). We also compared AutoQC with SAM-Only, i.e., SAM without prompt-feeding. It should be noted that since SAM-Only was incapable of classifying objects, we randomly assigned a category to each mask.

Figure 3a and 3b show representative segmentation results predicted by AutoQC (Fig. 3, a3 and b3), SAM-Only (Fig. 3, a4 and b4), and YOLOv8-Seg (Fig. 3, a5 and b5). Immunofluorescence and autofluorescence images were merged (Fig. 3, a1 and b1) to facilitate the identification of CMs and capillaries in the input images (Fig. 3, a2 and b2). SAM-Only demonstrated inaccurate segmentation, especially for small objects, and ineffective classification. YOLOv8-Seg showed better performance in segmentation and classification; however, it failed to detect all objects within input images. AutoQC demonstrated the best overall segmentation performance, although CMs and capillaries without clear edges presented a challenge. We evaluated segmentation performance using metrics mAP (Fig. 3c), mAR (Fig. 3d) and F1 score (Fig. 3e). SAM-Only had the lowest mAP, mAR, and F1 score across all IoU threshold conditions compared to YOLOv8-Seg and AutoQC (all $p$ values < 0.0001 compared to AutoQC). The evaluation results obtained from five test runs ($n$ = 5) are reported as mean ± SD in Table 1. AutoQC achieved the highest mAP, mAR, and F1 score, when computed using IoU threshold of 0.5 and IoU threshold range [0.5:0.95]. For the specific case, IoU threshold of 0.75, YOLOv8-Seg had a mAP of 0.701 and an F1 score of 0.717, with no significant difference compared to AutoQC; however, YOLOv8-Seg had a significantly lower mAR ($p$ = 0.0338).

We further evaluated model performance of capillarization assessment (Fig. 3, f-l) by calculating assessment errors (see **Performance evaluation of capillarization assessment**). SAM-Only had the highest counting errors (CMs, 0.480 ± 0.0468; capillaries, 1.11 ± 0.0498). AutoQC demonstrated the lowest counting errors (CMs, 0.122 ± 0.0131; capillaries, 0.098 ± 0.0062), which were significantly lower than those of YOLOv8-Seg (CMs, 0.243 ± 0.0298; capillaries, 0.124 ± 0.0131). In addition, AutoQC outperformed in measuring the total area of CMs, with an error of 0.067 ± 0.0115; the total area of capillaries, with an error of 0.157 ± 0.0299. We found that YOLOv8-Seg had no significantly different performance in measuring the total area of capillaries compared to AutoQC, with an error of 0.185 ± 0.0333. Furthermore, AutoQC demonstrated the lowest errors when assessing CDFA (0.0984 ± 0.0062) and CDCA (0.148 ± 0.0186). YOLOv8-Seg showed significantly higher errors when computing CDFA (0.124 ± 0.0131) and CDCA (0.240 ± 0.0288). In contrast, SAM-Only had dramatically higher CDFA and CDCA errors with values of 1.11 ± 0.0498 and 3.83 ± 0.5650, respectively. Similarly, when computing CCR, AutoQC achieved an error of 0.189 ± 0.0154, significantly lower than that of YOLOv8-Seg (0.351 ± 0.0423) and SAM-Only (0.870 ± 0.1010).

## Discussion

We proposed an automated image analysis method, AutoQC, to quantitatively assess cardiac capillarization in single-immunostained myocardial samples. We have demonstrated that AutoQC can effectively identify, classify, and segment CMs and capillaries in immunofluorescence images of basement membrane protein, collagen type IV.

To zero-shot transfer SAM to our task, i.e., segmenting CMs and capillaries in immunofluorescence images, we employed prompt engineering by enhancing bounding boxes with sets of binary-labeled centroids, serving as fine constraints. In the ablation study, we demonstrated this prompt design can effectively adapt SAM to microscopic bioimages and solve segmentation problems without the need for additional training. Prompting SAM with bounding boxes alone was found to yield better performance compared to using sets of binary-labeled centroids alone. This can be attributed to the fact that centroids, as the internal keypoints of objects, lack the information of object boundaries, although they define the relative location of target and adjacent objects. Keypoint-based detectors, such as detecting each object as a triplet with one center point and two corners[26] or five keypoints including extreme points and one center point[27], have reported to be effective in object detection. Hence, it is possible to improve P-SAM performance by optimizing the design of keypoints, potentially involving the implementation of keypoint detectors. Further study is needed to assess the performance improvement that optimizing keypoints may bring to AutoQC.

Although trained on a small dataset, AutoQC demonstrated impressive performance on instance segmentation of microscopic images. AutoQC achieved significantly higher mAP, mAR and F1 score compared to the state-of-the-art model YOLOv8-Seg. Furthermore, AutoQC showed promising performance in computing capillarization-related measurements, yielding CDFA error of 0.0984, CDCA error of 0.148, and CCR error of 0.189. In contrast, SAM-Only showed limitations on handling densely packed objects, leading to inaccurate segmentation. This finding highlights the importance of prompting for transferring SAM to new tasks, especially in challenging scenarios with densely packed objects. In addition, SAM-Only is unable to classify objects, making additional algorithms essential for classification purposes. It should be noted that the segmentation of small objects, capillaries in this study, is more challenging compared to that of large objects, CMs. By expanding the training dataset, AutoQC's instance segmentation performance on capillaries should be improved. Although it involves increased work of image acquisition and data annotation, the weak annotation process reduces the overall workload compared to pixel-wise annotations required by end-to-end segmentation models.

Confocal laser scanning microscopy allows 3D visualization of proteins of interest due to its optical-sectioning capability. By collecting a stack of images, AutoQC has the potential to analyze each individual image of the whole stack, allowing for 3D capillarization assessment. An imaging system with a high axial resolution and acquiring Z stacks with an appropriate Nyquist sampling interval are essential for maintaining the accuracy of 3D reconstruction and capillarization assessment. Hence, it is of great significance to relieve the burden of laborious data acquisition and labor-intensive data annotation by adopting a weak-supervised learning framework.

We demonstrated that CMs and capillaries can be simultaneously labeled by immunostaining of collagen type IV, and AutoQC can effectively compute capillarization-related measurements by automatically analyzing acquired immunofluorescence images. Alternatively, laminins, another major component of basement membrane, can be immunostained for detection of CMs and endothelial cells. It should be noted that only laminin isoforms expressed in both CM basement membrane and endothelial basement membrane should be considered, such as isoforms containing subunit α4 and/or γ1[28–30]. In general, AutoQC has the potential to analyze immunofluorescence images of structural proteins, as long as the protein distribution encloses both CMs and capillaries and can be efficiently detected using confocal fluorescence microscopy.

In summary, we have developed AutoQC, an automated image analysis tool for cardiac capillarization assessment of single-immunostained myocardial samples. CMs and capillaries in myocardial slices were labeled by immunofluorescence of a basement membrane protein, collagen type IV. Immunostained myocardial slices were examined using confocal fluorescence microscopy. AutoQC automatically classified and segmented each CM and capillary within input immunofluorescence images to compute capillarization-related measurements, including capillary density (CDFA and CDCA) and CCR. We have demonstrated that AutoQC outperformed the state-of-the-art instance segmentation model, YOLOv8-Seg. Moreover, AutoQC was trained on a small dataset without the need for pixel-wise annotation, reducing the workload associated with annotation and training. As such, AutoQC provides an efficient and practical solution for quantitative assessment of cardiac capillarization when performing single immunostaining on myocardial samples.

## Methods

### Slice preparation

All procedures on mice and animal care were approved by the Institutional Animal Care and Use Committee (IACUC) of the Medical University of South Carolina (protocol number, IACUC-2019-00868). All animal experiments were performed in accordance with ARRIVE guidelines and local regulations. C57BL/6 mice were placed in a euthanasia chamber, and euthanasia was conducted by gradually filling the chamber with 100% $CO_2$ for at least 3 min. Ventricles were isolated from C57BL/6 mouse hearts, fixed in freshly prepared 4% (w/v) paraformaldehyde, cryoprotected by sucrose solution, then snap-frozen in liquid-nitrogen-cooled isopentane. Frozen blocks were consecutively sectioned into 10 µm-thick slices and mounted on microscope slides (VWR International, Radnor, PA).

### Immunofluorescence

Heat-induced antigen retrieval was performed in sodium citrate pH 6.0 at 95 °C. Slices were blocked in 2% bovine serum albumin with 0.3% Triton X-100 at room temperature, then incubated with antibodies to collagen type IV (ab19808; Abcam, Boston, USA) at 4 °C. Slices were stained with abberior STAR RED (STRED; abberior, Göttingen, Germany) at room temperature. After staining, slices were mounted in custom-made Mowiol medium and coverslipped. Primary and secondary antibodies were diluted in the blocking buffer. Negative-primary-antibody controls were performed to confirm the staining signal was from the detection of target antigens.

### Home-built confocal fluorescence microscope

Confocal fluorescence imaging was performed on a home-built multimodal microscope[31]. Briefly, two picosecond pulsed lasers (470 nm and 635 nm; PicoQuant) were synchronized, serving as excitation. Two excitation beams were combined by dichroic mirrors (FF662-FDi01 and Di02-R488; Semrock), deflected by an XY two-axis galvanometer set (8310K; Cambridge Technology), expanded by a lens pair (89683, Edmund Optics; TTL200, Thorlabs), directed to the back aperture of a HCX Plan Apo 63x oil NA 1.40 objective (Leica Microsystems).

For the detection of abberior-STAR-RED-stained collagen type IV, the excitation laser at 635 nm was used. The emission light was spectrally separated from the excitation lights and green fluorescence by dichroic mirrors (Di02-R488, FF662-FDi01 and FF735-Di01; Semrock), filtered by a bandpass filter (690/50;

Chroma), coupled into a multimode fiber (M31L01; Thorlabs), and detected by a single photon detector (SPCM-AQRH-13; Excelitas Technologies Corp.). For the detection of green autofluorescence in formaldehyde-fixed myocardial slices, the excitation laser at 470 nm was used. The emission light was spectrally separated from the excitation lights and red fluorescence by dichroic mirrors (Di02-R488, FF662-FDi01 and FF624-Di01; Semrock), filtered by a bandpass filter (FF01-525/50; Semrock), coupled into a multimode fiber (M42L01; Thorlabs), and detected by a single photon detector (SPCM-AQRH-13; Excelitas Technologies Corp.).

**Image acquisition and dataset preparation**

Immunofluorescence and autofluorescence images were sequentially acquired in a frame-scan mode using SciScan (Scientifica, Uckfield, UK). Each region of interest (ROI) was imaged with a constant field of view (FOV), 42.5 μm × 42.5 μm of myocardial cross-sectioned area. A total of 107 ROIs were imaged. Raw images (512 × 512 pixels, 16-bit grayscale) were pre-processed by Wiener filtering (kernel size, 3×3) and contrast adjustment using a custom MATLAB script. Immunofluorescence images were annotated by two observers, reaching a consensus. Autofluorescent features in formaldehyde-fixed myocardial slices were used to facilitate the identification of CMs and capillaries in single-immunofluorescence images[32]. For the training of the object detector (YOLOv8-based) in the prompt-learning block, bounding box annotations with categories (CM or capillary) were utilized. For the training of the end-to-end instance segmentation model (YOLOv8-Seg), pixel-wise annotations with categories (CM or capillary) were utilized. Figure 4 summarizes the process of dataset preparation, annotation generation, and model training.

**AutoQC Algorithm Design**

*Overall architecture*

AutoQC harnesses the power of SAM, a robust foundation model for segmentation, and enables zero-shot transfer to our segmentation task by an automated prompt-learning block with classification capability (Fig. 5). The input image is a grayscale immunofluorescence image, where CMs and capillaries are outlined by immunostaining of collagen type IV. The input image is processed by the prompt-learning block to generate prompts and assign categories (CM or capillary) for each individual object. Together with the input image, the prompts and their associated categories are fed into the SAM-based segmentation block. The segmentation block takes input prompts and outputs pixel-level segmentation masks without the need for additional training. Commonly used capillarization-related measurements can then be computed from these segmentation results.

*Design of the prompt-learning block*

The prompt-learning block is designed for complete automation without the need for manual prompt crafting. The prompt-learning block consists of an object detector and a prompt generator. The YOLOv8-based object detector[22] is responsible for identifying, locating, and classifying objects, as well as predicting bounding boxes. An object detection task requires less annotated data for training compared to a segmentation task. In addition, bounding boxes are defined by corner points, allowing for lightweight storage and faster operation.

In scenarios where objects are sparsely distributed within images, the overlap of bounding boxes is negligible. Hence, directly employing bounding boxes as prompts can effectively zero- or few-shot transfer pre-trained models to custom datasets. However, biological structures typically organize in a dense manner and exhibit complicated protein patterns in microscopic bioimages. Within the myocardium, CMs are tightly packed, and each CM is surrounded by multiple capillaries. The predicted bounding boxes of CMs and capillaries showed pronounced overlap (Fig. 6). Due to the non-convex hull shape of CMs, the bounding boxes of adjacent CMs always overlapped with each other (Fig. 6b and 6c). In addition, the bounding boxes of capillaries were frequently enclosed by nearby CMs (Fig. 6e and 6f). We found that direct prompting SAM with these bounding boxes had an adverse effect on segmentation performance. As such, we designed a prompt generator to ensure bounding boxes appropriately function as prompts when adapting SAM to the custom dataset.

For a specific object in the input image, the bounding box $B_i$ works as a coarse constraint that can limit the maximum range of the segmentation mask and provide the segment-based category. Each bounding box $B_i$ has a centroid, $C_i$. To ensure a bounding box properly functions as a prompt, a set of binary-labeled points $S_i$ is generated as a fine constraint (Fig. 7). For the bounding box $B_i$ of a specific object, the point set $S_i$ is generated by traversing centroids of all recognized objects in the image and determining if they fall within the region of $B_i$. Centroids located within the region of $B_i$ are declared as the elements of $S_i$, with assigned binary labels. For all possible elements $C_j$ in the point set $S_i$, if $j \neq i$, $C_j$ is declared as $\{C_j, 0\}$; if $j =$

$i$, $C_j$ is declared as $\{C_j, 1\}$. This prompt-design approach requires a maximal complexity of $O(n^2)$ for segmenting an image, where $n$ refers to the number of recognized objects in the image.

*SAM-based segmentation block*

The SAM-based segmentation block consists of an image encoder, a prompt encoder, and a mask decoder. The image encoder for generating embeddings of the input image is based on a masked autoencoder[33]. The prompt encoder for creating embeddings of the prompts generated by the prompt-learning block is based on CLIP[34]. The mask decoder is based on a conventional transformer decoder[35], utilizing both image and prompt embeddings to predict segmentation masks. It should be noted that each prompt corresponds to a mask prediction. In this study, we utilized the pre-trained model weights of SAM and did not perform any tuning.

**Model setup**

All training, validation, and testing were executed on one node of the Palmetto cluster (Clemson University, Clemson, USA) with an Intel® Xeon® Platinum 8358 processor and an NVIDIA® A100 GPU. Two models, the YOLOv8-based object detector and YOLOv8-Seg, were trained using the stochastic gradient descent optimizer with a learning rate of 0.01. A momentum of 0.937 was incorporated into the training process, allowing for faster convergence and reduced oscillation. The mini-batch size was set to 16 and each model was trained for 100 epochs. We used 70% of the dataset for training and 30% for testing. 10% of the training dataset was used for validation. During training, images were augmented by random rotation and mosaic. The model weights of each model were optimized on the validation dataset.

**AutoQC Performance Analysis**

*Performance evaluation of instance segmentation*

COCO standard metrics[36] were utilized to evaluate segmentation performance. Mean average recall (mAR), mean average precision (mAP), and F1 score were computed across all categories. Intersection over Union (IoU) is commonly used to measure the degree of overlap between a predicted mask ($M_p$) and the ground truth ($M_{gt}$). When IoU ($M_p$, $M_{gt}$) ≥ $\phi$, the predicted mask with the identical category as the ground truth is considered a true positive (TP), where $\phi$ refers to a custom-defined IoU threshold. When IoU ($M_p$, $M_{gt}$) < $\phi$, the predicted mask with the identical category as the ground truth is considered a false positive (FP). If the category of the predicted mask does not match the ground truth, the predicted mask is considered a true negative (TN) when IoU ($M_p$, $M_{gt}$) ≥ $\phi$; a false negative (FN) when IoU ($M_p$, $M_{gt}$) < $\phi$. Specifically, we used two single-value IoU thresholds, 0.5 and 0.75; one IoU threshold range [0.5:0.95] with a step size of 0.05. When computing segmentation metrics under the range [0.5:0.95], results of each metric were averaged across all categories and ten IoU thresholds within the range. A detailed definition of the evaluation metrics used in this study is summarized in Table 2. It should be noted that accuracy was not considered in this study to avoid biased performance evaluation. For instance, given a scenario where 9 capillaries and a single CM exhibit in an image, if a model correctly predicts all capillaries but fails to predict the only CM, its accuracy of 90% is misleading and insufficient to reflect the performance in predicting CMs.

*Performance evaluation of capillarization assessment*

From each image, the area of each object mask was measured in $\mu m^2$, and the number of objects in each category (CM or capillary) was counted. The following capillarization-related measurements were computed for each image:

a. Total number of CMs, was counted as the number of masks with a CM category.

b. Total number of capillaries, was counted as the number of masks with a capillary category.

c. Total area of CMs ($\mu m^2$), was calculated as the area occupied by masks with a CM category.

d. Total area of capillaries ($\mu m^2$), was calculated as the area occupied by masks with a capillary category.

e. Capillary density in relation to FOV area (CDFA, $\mu m^{-2}$), was defined as capillaries per FOV area and calculated as the total number of masks with a capillary category divided by the FOV.

f. Capillary density in relation to CM area (CDCA, $\mu m^{-2}$), was defined as capillaries per CM area and calculated as the total number of masks with a capillary category divided by the total area of masks with a CM category.

g. Capillary-to-CM ratio (CCR), was calculated as the ratio of the number of masks with a capillary category to the number of masks with a CM category.

For each capillarization-related measurement, the assessment error was calculated as,

$$\delta = \frac{(predicted\ result) - (ground\ truth)}{ground\ truth}$$

**Statistics**

Five test runs were conducted to ensure fairness and reliability when comparing different models. During each test run, the training dataset was randomly shuffled before model setup (see **Model setup**).

Performance evaluation was conducted for each test run, and evaluation results ($n$ = 5) were reported as mean ± standard deviation (SD). Statistical differences were examined using unpaired or paired two-tailed *t* tests in Prism (GraphPad Software). A significance level of $p < 0.05$ was considered statistically significant.

## References

1. Hinkel, R. *et al.* Diabetes mellitus–induced microvascular destabilization in the myocardium. *J. Am. Coll. Cardiol.* **69**, 131–143 (2017).

2. Shimizu, I. *et al.* Excessive cardiac insulin signaling exacerbates systolic dysfunction induced by pressure overload in rodents. *J. Clin. Invest.* **120**, 1506–1514 (2010).

3. Kumamoto, H. *et al.* Beneficial effect of myocardial angiogenesis on cardiac remodeling process by amlodipine and MCI-154. *Am. J. Physiol. Circ. Physiol.* **276**, H1117–H1123 (1999).

4. Mohammed, S. F. *et al.* Coronary microvascular rarefaction and myocardial fibrosis in heart failure with preserved ejection fraction. *Circulation* **131**, 550–559 (2015).

5. Sano, M. *et al.* p53-induced inhibition of Hif-1 causes cardiac dysfunction during pressure overload. *Nature* **446**, 444–448 (2007).

6. Yoshizawa, S., Uto, K., Nishikawa, T., Hagiwara, N. & Oda, H. Histological features of endomyocardial biopsies in patients undergoing hemodialysis: Comparison with dilated cardiomyopathy and hypertensive heart disease. *Cardiovasc. Pathol.* **49**, 107256 (2020).

7. Rakusan, K., Flanagan, M. F., Geva, T., Southern, J. & Van Praagh, R. Morphometry of human coronary capillaries during normal growth and the effect of age in left ventricular pressure-overload hypertrophy. *Circulation* **86**, 38–46 (1992).

8. Iemitsu, M., Maeda, S., Jesmin, S., Otsuki, T. & Miyauchi, T. Exercise training improves aging-induced downregulation of VEGF angiogenic signaling cascade in hearts. *Am. J. Physiol. Heart Circ. Physiol.* **291**, H1290-8 (2006).

9. Roh, J. D. *et al.* Exercise training reverses cardiac aging phenotypes associated with heart failure with preserved ejection fraction in male mice. *Aging Cell* **19**, e13159 (2020).

10. Larouche, I. & Schiffrin, E. L. Cardiac microvasculature in DOCA-salt hypertensive rats : Effect of endothelin ET(A) receptor antagonism. *Hypertension* **34**, 795–801 (1999).

11. Miyachi, M. *et al.* Exercise training alters left ventricular geometry and attenuates heart failure in Dahl salt-sensitive hypertensive rats. *Hypertension* **53**, 701–707 (2009).

12. Battistella, R. *et al.* Not all lectins are equally suitable for labeling rodent vasculature. *Int. J. Mol. Sci.* **22**, (2021).

13. Emde, B., Heinen, A., Gödecke, A. & Bottermann, K. Wheat germ agglutinin staining as a suitable method for detection and quantification of fibrosis in cardiac tissue after myocardial infarction. *Eur. J. Histochem.* **58**, 2448 (2014).

14. Jilani, S. M. *et al.* Selective binding of lectins to embryonic chicken vasculature. *J. Histochem. Cytochem. Off. J. Histochem. Soc.* **51**, 597–604 (2003).

15. Johansson, B., Mörner, S., Waldenström, A. & Stål, P. Myocardial capillary supply is limited in hypertrophic cardiomyopathy: A morphological analysis. *Int. J. Cardiol.* **126**, 252–257 (2008).

16. Silvestre, J.-S. *et al.* Proangiogenic effect of angiotensin-converting enzyme inhibition is mediated by the bradykinin B2 receptor pathway. *Circ. Res.* **89**, 678–683 (2001).

17. Heinrich, L. *et al.* Whole-cell organelle segmentation in volume electron microscopy. *Nature* **599**, 141–146 (2021).

18. Rivenson, Y. *et al.* Virtual histological staining of unlabelled tissue-autofluorescence images via deep learning. *Nat. Biomed. Eng.* **3**, 466–477 (2019).

19. Greenwald, N. F. *et al.* Whole-cell segmentation of tissue images with human-level performance using large-scale data annotation and deep learning. *Nat. Biotechnol.* **40**, 555–565 (2022).

20. He, K., Gkioxari, G., Dollár, P. & Girshick, R. Mask r-cnn. In *Proc. of the IEEE Int. Conf. Comput. Vis.* 2961–2969 (2017).

21. Liu, S., Qi, L., Qin, H., Shi, J. & Jia, J. Path aggregation network for instance segmentation. In *Proc. Of the IEEE Conf. Comput. Vis. Pattern Recognition* 8759–8768 (2018).

22. Jocher, G., Chaurasia, A. & Qiu, J. YOLO by ultralytics. https://github.com/ultralytics/ultralytics/blob/main/CITATION.cff (2023).

23. Liu, P. *et al.* Pre-train, prompt, and predict: A systematic survey of prompting methods in natural language processing. *ACM Comput. Surv.* **55**, (2023).

24. Wang, J. *et al.* Review of large vision models and visual prompt engineering. *arXiv Prepr.* arXiv2307.00855 (2023).

25. Kirillov, A. *et al.* Segment anything. *arXiv Prepr.* arXiv2304.02643 (2023).

26. Duan, K. *et al.* Centernet: Keypoint triplets for object detection. In *Proc. of the IEEE/CVF Int. Conf. Comput. Vis.* 6569–6578 (2019).

27. Zhou, X., Zhuo, J. & Krahenbuhl, P. Bottom-up object detection by grouping extreme and center points. In *Proc. of the IEEE/CVF Conf. Comput. Vis. Pattern Recognition* 850–859 (2019).

28. Iivanainen, A., Sainio, K., Sariola, H. & Tryggvason, K. Primary structure and expression of a novel human laminin alpha 4 chain. *FEBS Lett.* **365**, 183–188 (1995).

29. Iivanainen, A. *et al.* Primary structure, developmental expression, and immunolocalization of the murine laminin α4 chain. *J. Biol. Chem.* **272**, 27862–27868 (1997).

30. Granath, C. *et al.* Characterization of laminins in healthy human aortic valves and a modified decellularized rat scaffold. *Biores. Open Access* **9**, 269–278 (2020).

31. Zhang, Z. *et al.* Multimodal microscopy imaging of cardiac collagen network: Are we looking at the same structures? In *Proc.SPIE on Diagnostic and Therapeutic Applications of Light in Cardiology 2023* **12355**, 52-56 (2023).

32. Zhang, Z. *et al.* Management of autofluorescence in formaldehyde-fixed myocardium: Choosing the right treatment. *Eur. J. Histochem.* **67**, 3812 (2023).

33. He, K. *et al.* Masked autoencoders are scalable vision learners. In *Proc. of the IEEE/CVF Conf. Comput. Vis. Pattern Recognition* 16000–16009 (2022).

34. Radford, A. *et al.* Learning transferable visual models from natural language supervision. In *Int. Conf. on Mach. Learn.* 8748–8763 (2021).

35. Vaswani, A. *et al.* Attention is all you need. *Adv. Neural Inf. Process. Syst.* **30** (2017).

36. Lin, T.-Y. *et al.* Microsoft coco: Common objects in context. In *Proc. European Conf. on Comput. Vis.* **8693**, 740–755 (2014).

## Acknowledgements

This research was supported by South Carolina IDeA Networks of Biomedical Research Excellence (SC INBRE), a National Institutes of Health (NIH) funded center (Award P20 GM103499), MTF Biologics Extramural Research Grant, South Carolina Translation Research Improving Musculoskeletal Health (TRIMH), an NIH funded Center of Biomedical Research Excellence (Award P20 GM121342), and an R01 research grant to BG (R01HL144927). This work was supported in part by the National Science Foundation EPSCoR Program under NSF Award #OIA-1655740 and OIA-2242812.

## Author contributions

Z. Z. conceived and designed the experiments, collected and annotated the data, evaluated models' capillarization assessment performance, wrote the main text of the manuscript, prepared all figures and tables. X. C. annotated the data, designed AutoQC, evaluated models' instance segmentation performance, wrote the draft of AutoQC algorithm design and model setup. T. Y. helped draft the manuscript. W. R., B. G., A. R. and T. Y. provided secured funding, supervised the project, and contributed to manuscript revision. All authors reviewed the manuscript before publication.

## Data availability

Image dataset and algorithm code are available from the corresponding author upon request.

## Competing interests

The authors declare no competing interests.

**Figures and legends**

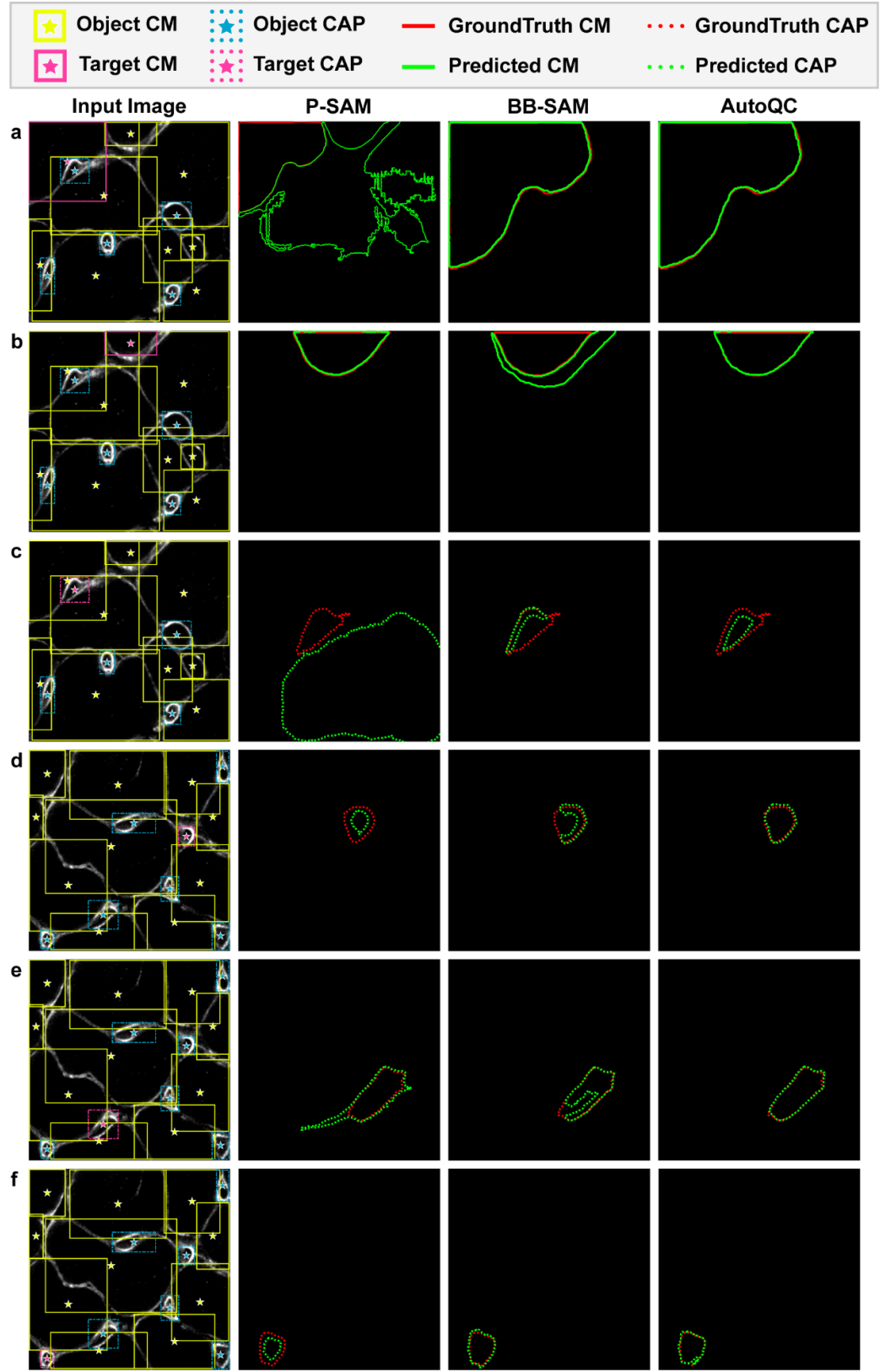

Figure 1. Representative segmentation results predicted by P-SAM, BB-SAM, and AutoQC in the ablation study. Segmentation results of two images from different test runs are presented. For each input image, the predicted masks (green) of three target objects, either CM or capillary, are displayed and overlapped with the ground truth masks(red). The bounding box and its centroid of each target object are highlighted (pink) in the input image. CM, objects with a cardiomyocyte category denoted by solid line; CAP, objects with a capillary category denoted by dashed line.

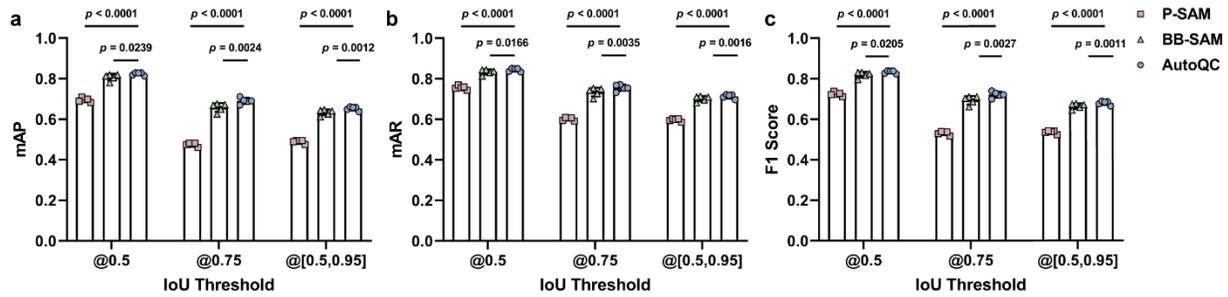

Figure 2. Quantitative evaluation of instance segmentation among P-SAM, BB-SAM, and AutoQC in the ablation study. **(a)** mAP, **(b)** mAR, and **(c)** F1 score were computed using IoU threshold 0.5, 0.75, and [0.5:0.95] with a step of 0.05. Paired two-tailed *t* tests were used to compare evaluation results between AutoQC and P-SAM or BB-SAM. *p* < 0.05 was considered significantly different.

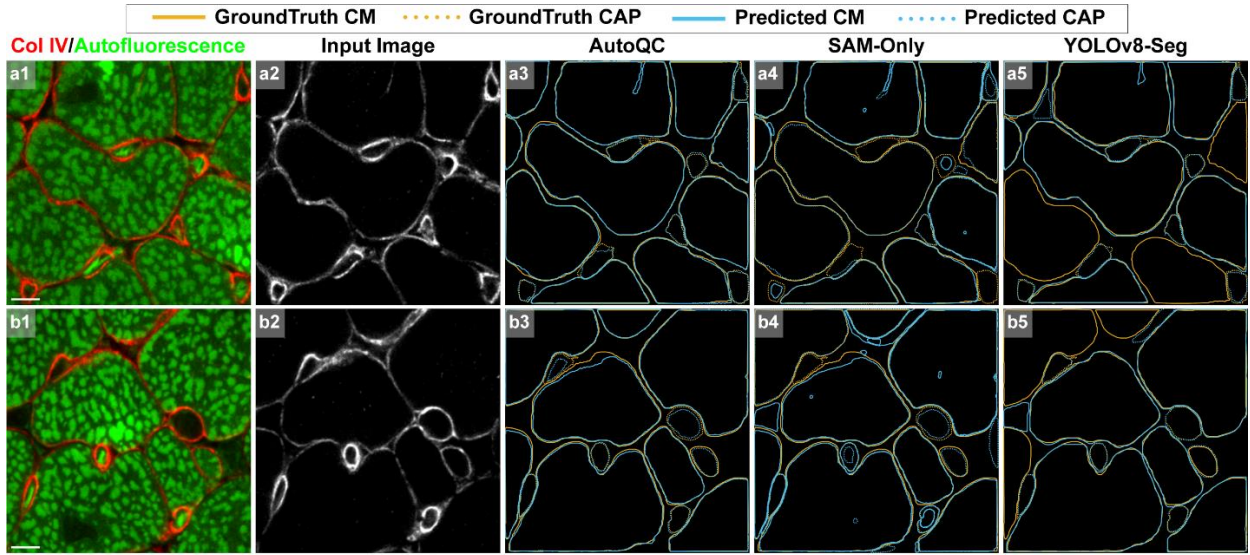
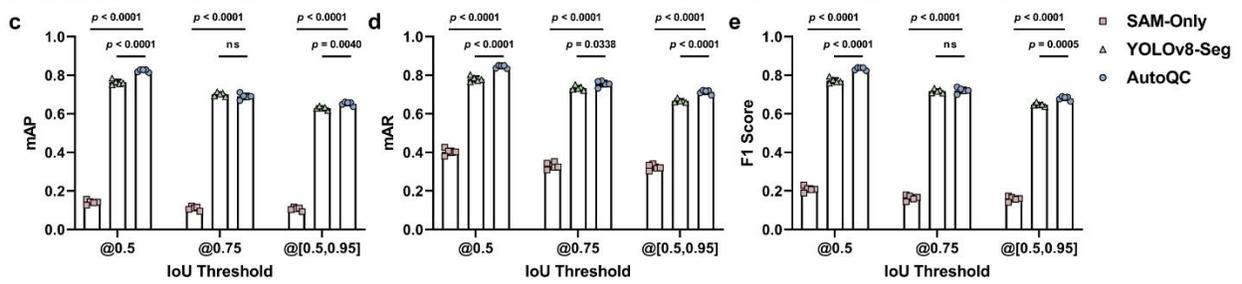
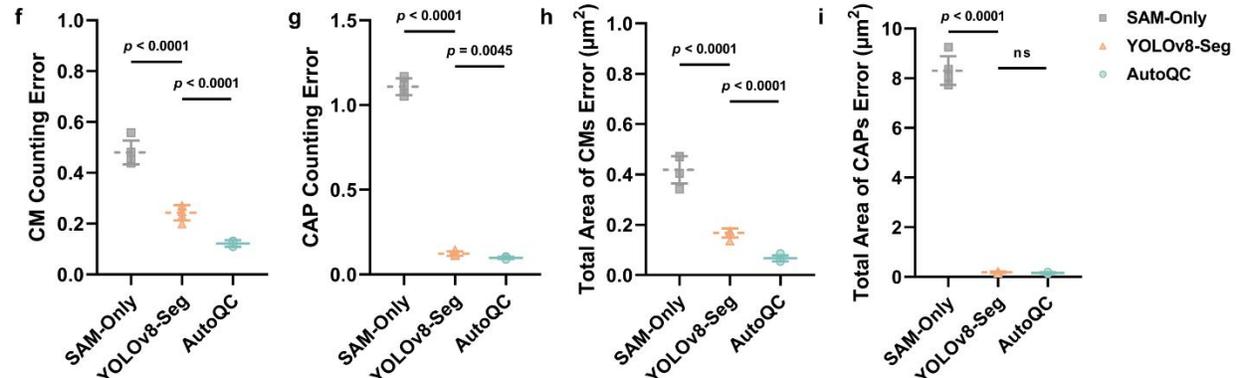
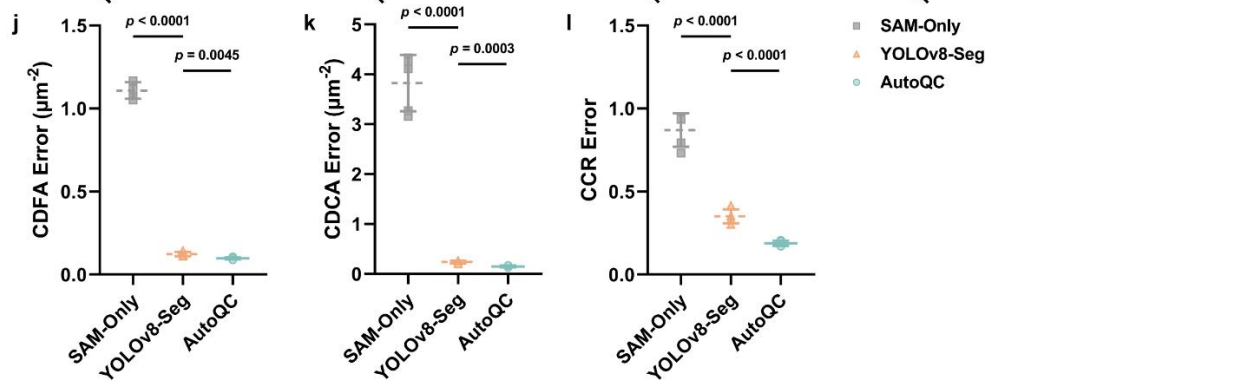

Figure 3. Performance evaluation among AutoQC, SAM-Only, andYOLOv8-Seg. **(a, b)** Representative segmentation results of two images from different test runs. (a1, b1) Immunofluorescence images of collagen type IV (ColIV, red) are merged with autofluorescence images (green) to facilitate the interpretation of CMs and capillaries in (a2, b2) input images. Predicted masks output by (a3, b3) AutoQC, (a4, b4) SAM-Only, and (a5, b5) YOLOv8-Seg are overlap with the ground truth. CM, objects with a cardiomyocyte category denoted by solid line; CAP, objects with a capillary category denoted by dashed line; Scale bars, 5 μm. **(c)** mAP, **(d)** mAR, and **(e)** F1 score were computed using IoU threshold 0.5, 0.75, and [0.5:0.95] with a step of 0.05. **(f)** CM counting, **(g)** capillary counting, **(h)** total area of CMs, **(i)** total area of capillaries, **(j)** CDFA, **(k)** CDCA, and **(l)** CCR are presented by the assessment errors between predicted results and the ground truth. Error bars represent the standard deviation of assessment errors in five test runs; unpaired two-tailed *t* tests; $p < 0.05$ was considered significantly different.

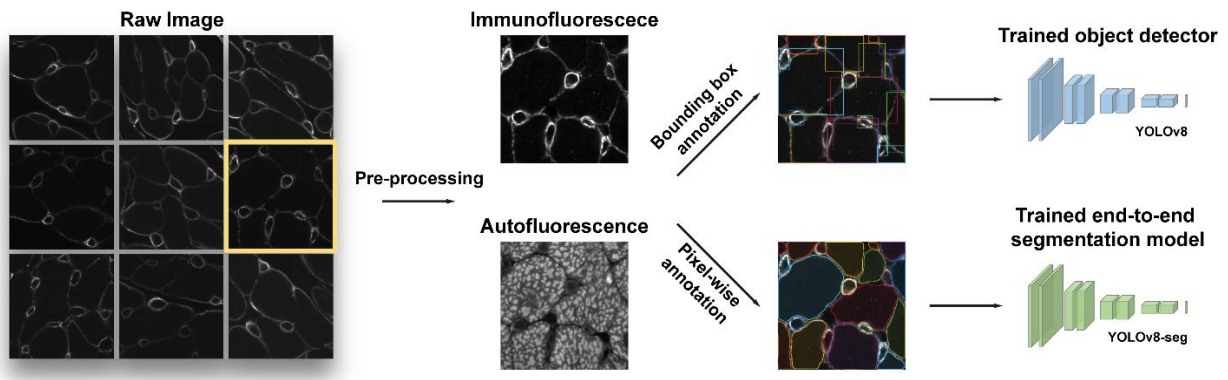

Figure 4. The model training workflow, including image pre-processing, utilizing autofluorescence for identifying and annotating CMs and capillaries in immunofluorescence images of collagen type IV, weak annotations (bounding boxes) for training the object detector, and pixel-wise annotations for training the end-to-end instance segmentation model.

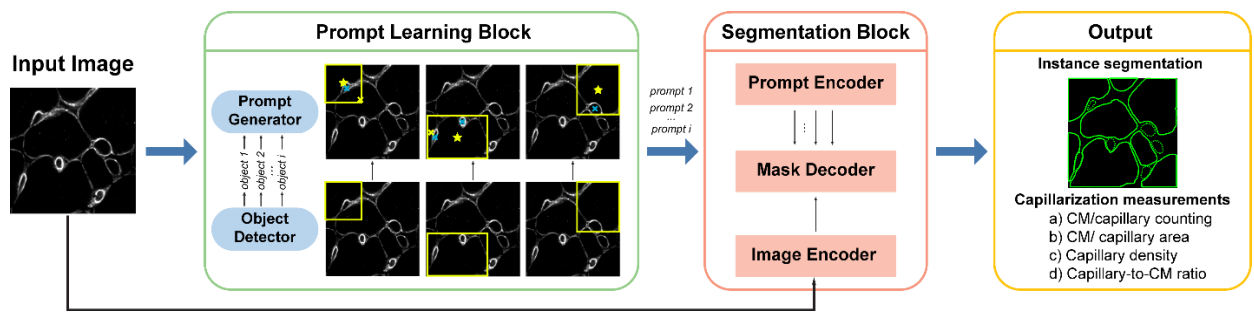

Figure 5. Overview of AutoQC algorithm. An input image is fed into both the prompt-learning block and the SAM-based segmentation block. The prompt-learning block is responsible for generating prompts and classifying objects. The segmentation block takes input prompts from the prompt-learning block to segment objects in the input image without the need for additional training. Commonly used capillarization-related measurements can be computed from the segmentation results.

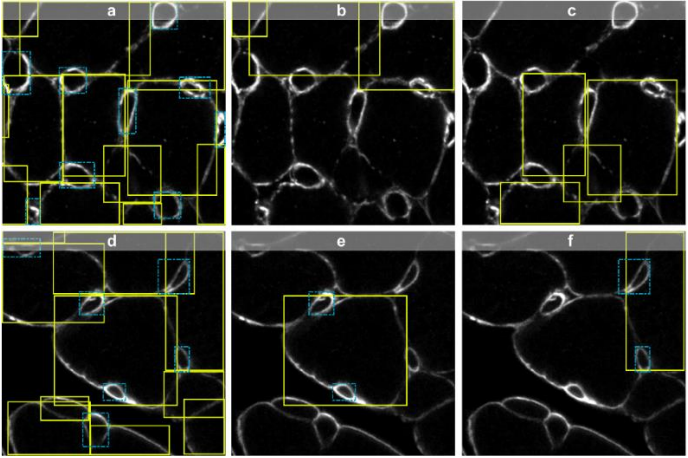

Figure 6. Representative immunofluorescence images of densely packed CMs and capillaries. Bounding boxes of CMs are marked by solid lines (yellow), and those of capillaries are marked by dashed lines (blue). (a, d) Densely packed CMs and capillaries result in largely overlapped bounding boxes. (b, c) Pronounced overlap among the bounding boxes of adjacent CMs. (e, f) Some bounding boxes of capillaries are fully enclosed within the bounding boxes of nearby CMs.

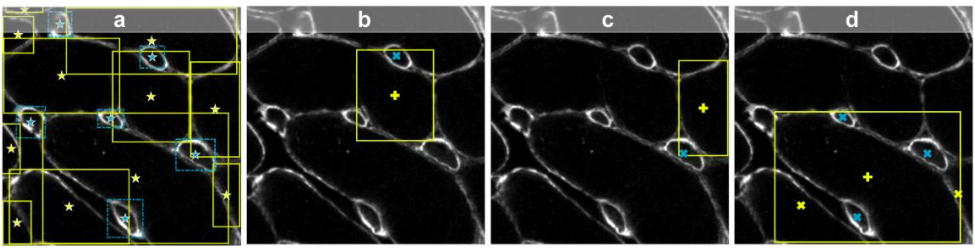

Figure 7. (a) Representative input image with bounding boxes of CMs marked by solid lines (yellow), and those of capillaries marked by dashed lines (blue). The centroids of bounding boxes are indicated by a pentagram sign with a matching color. (b-d) The prompt for a target object consists of a bounding box $B_i$ and a set of binary-labeled points $S_i$. The set of binary-labeled points $S_i$ includes: the centroid of the target bounding box ($j = i$) declared as $\{C_j, 1\}$ (plus sign); the centroids of other bounding boxes ($j \neq i$) that locate within the region of the target bounding box declared as $\{C_j, 0\}$ (cross sign).

## Tables

Table 1. Summary of instance segmentation evaluation results, reported as mean ± SD (five test runs).

|                | SAM-Only       | P-SAM          | BB-SAM         | YOLOv8-Seg     | AutoQC         |
|----------------|----------------|----------------|----------------|----------------|----------------|
| mAP @0.5       | 0.141±0.0111   | 0.695±0.0100   | 0.807±0.0165   | 0.764±0.0132   | *0.824±0.0070* |
| mAP @0.75      | 0.109±0.0101   | 0.475±0.0090   | 0.658±0.0211   | 0.701±0.0099   | *0.690±0.0148* |
| mAP @[0.5:0.95]| 0.106±0.0090   | 0.489±0.0081   | 0.634±0.0146   | 0.630±0.0082   | *0.653±0.0098* |
| mAR @0.5       | 0.403±0.0166   | 0.756±0.0096   | 0.834±0.0130   | 0.781±0.0145   | *0.844±0.0086* |
| mAR @0.75      | 0.331±0.0175   | 0.602±0.0080   | 0.734±0.0212   | 0.734±0.0117   | *0.756±0.0153* |
| mAR @[0.5:0.95]| 0.323±0.0147   | 0.596±0.0073   | 0.702±0.0139   | 0.667±0.0091   | *0.713±0.0107* |
| F1 @0.5        | 0.209±0.0144   | 0.724±0.0097   | 0.820±0.0148   | 0.772±0.0137   | *0.834±0.0078* |
| F1 @0.75       | 0.164±0.0135   | 0.531±0.0086   | 0.694±0.0212   | 0.717±0.0105   | *0.721±0.0148* |
| F1 @[0.5:0.95] | 0.159±0.0120   | 0.537±0.0078   | 0.665±0.0142   | 0.647±0.0083   | *0.680±0.0103* |

Table 2. Segmentation evaluation metrics used in the study. (*N* refers to the number of images)

| Metrics | Equations |
|---|---|
| Intersection over Union (IoU) | $IoU = \dfrac{M_p \cap M_{gt}}{M_p \cup M_{gt}}$ |
| Mean Average Recall (mAR) | $mAR = 1/N \sum_{i=1}^{N} AR_i$ |
| Mean Average Precision (mAP) | $mAP = 1/N \sum_{i=1}^{N} AP_i$ |
| F1 Score | $F1\ Score = 2 \times \dfrac{mAR \times mAP}{mAR + mAP}$ |